\documentclass[times,tighten,twocolumn,twocolappendix]{aastex631}

\newcommand{\GP}{{\scshape GALPROP}}
\newcommand{\hess}{{H.E.S.S.}}
\newcommand{\tibet}{Tibet~AS${\gamma}$}
\newcommand{\argo}{ARGO--YBJ}
\newcommand{\fermi}{\textit{Fermi}--LAT}
\newcommand{\graya}{$\gamma$-ray} 
\newcommand{\grays}{$\gamma$~rays} 

\shorttitle{Temporal Variability of Galactic TeV Emissions}
\shortauthors{Marinos et al.}

\begin{document}

\title{On the Temporal Variability of the Galactic Multi-TeV Interstellar Emissions}

\author[0000-0003-1734-0215]{P. D. Marinos}
\affiliation{W. W. Hansen Experimental Physics Laboratory and Kavli Institute for Particle Astrophysics and Cosmology, Stanford University, Stanford, CA 94305, USA}
\email{pmarinos@stanford.edu}
\affiliation{School of Physical Sciences, University of Adelaide, Adelaide, South Australia 5000, Australia}

\author[0000-0002-2621-4440]{T. A. Porter}
\affiliation{W. W. Hansen Experimental Physics Laboratory and Kavli Institute for Particle Astrophysics and Cosmology, Stanford University, Stanford, CA 94305, USA}

\author[0000-0002-9516-1581]{G. P. Rowell}
\affiliation{School of Physical Sciences, University of Adelaide, Adelaide, South Australia 5000, Australia}

\author[0000-0003-1458-7036]{G. Jóhannesson}
\affiliation{Science Institute, University of Iceland, IS-107 Reykjavik, Iceland}

\author[0000-0001-6141-458X]{I. V. Moskalenko}
\affiliation{W. W. Hansen Experimental Physics Laboratory and Kavli Institute for Particle Astrophysics and Cosmology, Stanford University, Stanford, CA 94305, USA}



\begin{abstract}
  %
  We use the GALPROP cosmic ray (CR) framework to model the Galactic CR distributions and associated non-thermal diffuse emissions up to PeV energies.
  We consider ensembles of discrete, finite lifetime CR sources, e.g.~supernova remnants (SNRs), for a range of creation rates and lifetimes.
  We find that global properties of the CR sources are likely not directly recoverable from the current `snapshot' of the historic injection and propagation of CRs within the Galaxy that are provided by the data.
  We show that models for the diffuse \grays{} based on the discrete/time-dependent scenarios we consider are able to explain the LHAASO very-/ultra-high energy (VHE/UHE) \graya{} data with up to 50\% contribution by unresolved leptonic sources at the highest energies.
  Over the models that we consider, variations in the diffuse VHE emissions can be $\sim$25\%, which is comparable to those for steady-state models that we investigated in earlier work.
  Such variations due to the discrete/finite nature of the CR sources are an important factor that are necessary to construct accurate physical models of the diffuse emissions from the Galaxy at VHE/UHEs.
\end{abstract}

\keywords{Particle Astrophysics (96); Cosmic Rays (329); Diffuse radiation (383); Gamma-rays (637); Interstellar Emissions (840)}



\section{Introduction}

Cosmic ray particles are thought to be accelerated up to at least PeV energies by sources within the Milky Way (MW). The CRs propagate for millions of years through the interstellar medium (ISM), resulting in a diffuse `sea' of particles from GeV energies up to the so-called CR `knee'.
As CRs diffuse they interact with various components of the ISM; the interstellar gas, the interstellar radiation field~(ISRF), and the Galactic magnetic field~(GMF), emitting non-thermal broad-band emissions from X-rays to PeV \grays{} in the process.
Observations of the very-high-energy~(VHE; $>$100\,GeV) \graya{} emission is essential in understanding how CRs are accelerated up to PeV energies and the mechanisms behind their propagation through the MW. Observations of VHE \grays{} also provides insights to the spatial distributions of the ISM components.

The diffuse \grays{} produced by the CR interactions in the ISM have been detected with high statistics in the 100\,MeV to 100s of GeV energy range by the \fermi~\citep[e.g.][]{2012ApJ...750....3A}, and \hess{} and \argo{} at TeV energies~\citep{2014PhRvD..90l2007A,2015ApJ...806...20B}. 
While interstellar emissions models~(IEMs) broadly agree with the sub-TeV observations~\citep[e.g.][]{2016ApJS..223...26A,2016ApJS..224....8A,2016ApJ...819...44A,2023MNRAS.518.5036M}, recent results from LHAASO and \tibet{} in the $>$10\,TeV regime indicates that there is an excess of \graya{} emission from the Galactic plane compared to their model predictions~\citep{2021ApJ...919...93F,2023PhRvL.131o1001C}.
This excess emission could be due to some contribution from \graya{} sources that are not individually resolved, or some other component of the ISM that is not well understood.
Connecting the diffuse \graya{} emission across this broad energy range is crucial for developing physical models for the emissions themselves.
These improved models will be required to extract properties from the faintest \graya{} sources, aiding in understanding how CRs are accelerated to PeV energies in the MW.

The often-used paradigm considered when modelling the diffuse emissions from the MW is the so-called `steady-state' approximation where CR sources follow continuous spatial distributions with constant injection rate, and the CR solution is obtained without any time-dependence \citep{1990acr..book.....B}.
Our previous work quantified the steady-state diffuse \graya{} modelling uncertainty up to 100\,TeV~\citep{2023MNRAS.518.5036M} and demonstrated a non-negligible fraction of leptonic emission above 1\,TeV.
However, the short cooling distances~($\lesssim$1\,kpc) of the $\gtrsim$10\,TeV electrons necessitates the consideration of individual sites of CR injection.
As the CR injection spectra are not known for every accelerator, and as distance estimates for many CR sources are on the order of kiloparsecs, there is some intrinsic variance in the models.
Such variance impacts all VHE emission predictions and has not yet been quantified.

In this paper, we take the next step and consider the effects by discrete/finite lifetime source ensembles, using the \GP\footnote{\url{http://galprop.stanford.edu}} framework to model the time-dependent CR solutions and associated non-thermal emissions from GeV to PeV energies. This extends the earlier work by \citep{2019ApJ...887..250P} to a broader range of model configurations, as well as a larger energy range.
The configurations that we consider place CR accelerators stochastically in space and time according to likely viable encompassing models for the large-scale source distributions in the MW. We consider that the sources can have individual lifetimes from 10\,kyr to 200\,kyr. Creation rates over the range of one source every 50\,yr to 500\,yr are also tested. For the CR electrons these variations allow us to fully explore the significant fluctuations in spectra due to these parameters \citep{2018JCAP...11..045M}.
We discuss the challenges in constraining the CR source parameters, namely the source lifetimes and creation rates, through observations of the diffuse \graya{} emission.
We also apply our modelling results to observations of the diffuse emission from the H.E.S.S. Galactic Plane Survey (HGPS)\footnote{\url{https://www.mpi-hd.mpg.de/hfm/HESS/hgps/}} and LHAASO in the energy range of 10\,TeV to 1\,PeV. We show that LHAASO's observed \graya{} excess can be explained naturally by a component of unresolved PWNe and discuss the potential to constrain the source parameters from the diffuse emission.

\section{Model Setup}

The \GP{} framework~\citep{1998ApJ...509..212S,1998ApJ...493..694M} is a widely employed CR propagation package with an extensive history of reproducing the local CR results, Galactic synchrotron emission, and the Galactic diffuse \graya{} emission~\citep[e.g.][]{2000ApJ...537..763S,2002ApJ...565..280M,2004ApJ...613..962S,2008ApJ...682..400P,2011A&A...534A..54S,2020ApJS..250...27B,2023MNRAS.518.5036M}. We use the latest release~(version~57), where an extensive description of the current features is given by~\citet{2022ApJS..262...30P}. The full configuration files can be found in the online supplemental material.

We choose ISM distributions with spiral-arm structures that are consistent with the data.
To place the CR sources we use a distribution with disc-like and spiral-arm components, with an equal relative contribution from each. This source distribution is hereafter referred to as SA50, where the number represents the percentage contribution from the spiral-arm component \citep[see][and references therein]{2017ApJ...846...67P}.
For the IC emission and pair-absorption calculations we use the ISRF developed by \citet{2017ApJ...846...67P} that includes spiral arms \citep[designated R12;][]{2012A&A...545A..39R}. This ISRF accounts for the anisotropic scattering cross section for the IC emission.
For the GMF we use the bisymmetric spiral model developed by \citet{2011ApJ...738..192P}, hereafter referred to as PBSS. This GMF model includes the large-scale regular and small-scale random components of the magnetic field.
For the 3D gas density we use the model developed by \citet{2018ApJ...856...45J} and references therein.
We take the CR injection and diffusion parameters from \citet{2023MNRAS.518.5036M} with their values included within the configuration files supplied online.
The CR injection and diffusion parameters are obtained following the procedure in \citet{2017ApJ...846...67P} and \citet{2018ApJ...856...45J} -- the post-diffusion spectra are fit to data from AMS--02 and \textit{Voyager~1} using the source distribution and ISM gas model \citep[see][and references therein]{2019ApJ...879...91J}.
As we normalise the CR spectra at energies above which the Solar modulation is critical, we use the force-field approximation instead of the more advanced \GP{}/HelMod framework \citep{2020ApJS..250...27B}.
For the CR diffusion calculations we use a non-linear spatial grid~\citep[tan spatial grid;][]{2022ApJS..262...30P} with the spatial grid size around the solar location set to 7\,pc.
We use ten kinetic energy bins per decade ranging from 1\,GeV\,nuc$^{-1}$ to 10\,PeV\,nuc$^{-1}$ for the CR propagation, and five bins per decade ranging from 1\,GeV to 1\,PeV for the \graya{} flux calculation.
The \graya{} skymaps utilise a seventh-order HEALPix~\citep{2005ApJ...622..759G} isopixelisation, giving a pixel size of $27.5^{\prime} \times 27.5^{\prime}$.
For the IC calculations we use the anisotropic scattering cross section~\citep{2000ApJ...528..357M}, and we account for the $\gamma \gamma \rightarrow e^{\pm}$ attenuation of the \grays{} on the ISRF above $\sim$10\,TeV using the formalism from \citet{2018PhRvD..98d1302P}.
Descriptions for the optimisation of the diffusion parameters can be found in \citet{2023MNRAS.518.5036M}, and references therein.

Comparisons between local hadronic CR observations and the steady-state \GP{} predictions are provided in~\citet{2017ApJ...846...67P} up to 1\,TeV\,nuc$^{-1}$. 
Comparisons between the local electron flux observations and the time-dependent \GP{} predictions will be discussed later in \autoref{sect:CRs}.

\subsection{Time-dependent CR source parameter optimisation} \label{ssect:t-dep parameters}

For the time-dependent solution we do not define the individual CR accelerator source types, e.g.~SNRs, PWNe, stellar clusters, and binary sources, as the relative CR contribution between the various source types in the MW is not adequately constrained.
Instead, we approximate some `average' source class distribution based on Galactic SNR/PWN models (see SA50, discussed above).
The source parameters are tuned under the steady-state assumption to reproduce the local CR spectra after propagation.
The injection spectra therefore represent an average over all potential sources that contribute to the local flux.
These spectra are then applied to each individual source in the time-dependent calculation.
The parameters for the CR injection spectra, and their values for the SA50 source distribution, can be found in Table~1 of \citet{2023MNRAS.518.5036M}.

When operating with the time-dependent solution we start by setting the CR density to zero across the entire MW.
Sources are then placed stochastically in position and time, with the CR sea density increasing until the injection and energy losses are balanced.
It is computationally expensive to perform these calculations for all CR species.
However, as previous results have illustrated that the $\pi^{0}$-decay emission is approximately constant over a 5\,Myr period~\citep{2019ApJ...887..250P}, we simplify the hadronic calculations to utilise the steady-state approximation (see~\autoref{sect:hadron results}).
For the leptonic emission we split the time-dependent propagation calculations into two periods: initial and final.
The initial period runs for 100\,Myr with a timestep of $\Delta t=2.5$\,kyr and allows the leptonic CR density to reach equilibrium.
The final propagation period continues the calculations for another 5\,Myr with the timestep decreased to $\Delta t=100$\,yr to ensure that the most rapid energy losses are accurately captured.
All analyses are performed only on the final 5\,Myr, with the CR electron densities and IC \graya{} flux being output once every 25\,kyr.

For the time-dependent solution, the creation rate and lifetime of the CR sources are additional free parameters which do not impact the CR normalisation condition but will impact the the energy-dependent magnitude of temporal variation.

Estimates on the rate of SNe vary depending on the observational technique. For example, estimates from the radioactive decay of $^{26}$Al range between one SNe every 35--125\,yr~\citep{2006Natur.439...45D}. Averaging over many observational techniques provides a tighter constraint of one SNe every 45--85\,yr~\citep{2021NewA...8301498R}. Statistical modelling from~\citet{2018JCAP...11..045M} suggests a lower rate of one SN every 500\,yr to explain the CR electron spectrum above 1\,TeV. Some unknown fraction of these SNRs will also have an associated PWNe.

Similarly, there is no tight constraint for the source lifetimes of the various source classes.
Estimates for the lifetimes of SNRs vary between 30--300\,kyr~\citep{2012JCAP...01..010B}. PWNe can accelerate CRs for longer than 100\,kyr, though the older PWNe contribute less to the Galactic CR electron flux density across all energies\footnote{The injection spectrum for the sources can be set to decay over their lifetimes. However, we do not use this function in the present paper.}~\citep{2020A&A...636A.113G}. Star clusters are another candidate for the acceleration of CRs up to PeV energies~\citep{2014A&ARv..22...77B}, and may accelerate CRs for up to 10\,Myr~\citep{2019NatAs...3..561A,2019RLSFN..30S.155B}, but are expected to be hadronic accelerators.
The presence of the \textit{Fermi} bubbles suggests that the GC region may experience phases of increased activity~\citep{2014ApJ...793...64A}; however, we do not model the GC as an additional variable source due to the CR power injected within being a free modelling parameter~\citep{2019ApJ...887..250P}.

\begin{table}
    \centering
    \caption{Names and source parameter values for the six tested combinations. Source lifetimes are given in units of kyr, and the source creation intervals are given as the average number of years between sources being created in the MW.}
    \label{tab:source pars}
    \begin{tabular}{ccc}
        \hline
        Combination Name & Lifetime~(kyr) & Creation Interval~(yr) \\
        \hline
        L010R100 & 10  & 100 \\
        L050R100 & 50  & 100 \\
        L100R100 & 100 & 100 \\
        L200R100 & 200 & 100 \\
        L100R500 & 100 & 500 \\
        L100R050 & 100 & 50  \\
        \hline
    \end{tabular}
\end{table}

To test the uncertainty in the time-dependent source parameters we test over the full range of values for the source parameters.
The source lifetime is varied from 10\,kyr up to 200\,kyr, and the source creation interval is varied from one source every 50\,yr to one source every 500\,yr~(i.e.~a source creation rate of 0.02--0.002\,yr$^{-1}$).
The source parameter combinations are denoted LxxxRyyy, where xxx represents the lifetime of the sources in kiloyears and yyy represents the creation interval in the average number of years between source creation.
Due to the computational expense of the calculations, we choose only six combinations for the source lifetimes and source creation intervals for the leptonic injectors.
As the L100R100 parameter combination uses the most commonly found values it is taken as our reference model. 
All six of our chosen parameter combinations can be found in \autoref{tab:source pars}.
As we are sampling at intervals of 25\,kyr and as the flux is expected to vary on timescales similar to the source lifetimes, the Nyquist condition is satisfied for all source parameter combinations except for L010R100.

\section{Cosmic-Ray Variability} \label{sect:CRs}

\begin{figure}
    \centering
    \includegraphics[width=8.5cm]{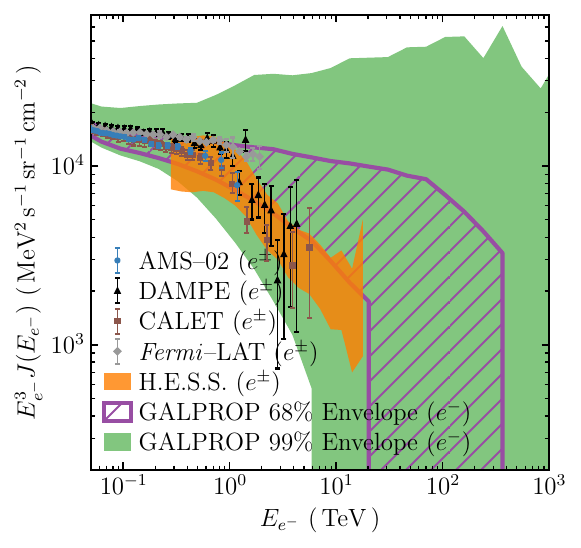}
    \caption{Kinetic energy spectrum for the primary CR electrons above 20\,GeV at the Solar location for the final 5\,Myr. The 68\% and 99\% envelopes of the electron spectrum over all six source parameter combinations~(\autoref{tab:source pars}) are shown by the purple hatched area and green shaded band, respectively. The combined electron and positron spectrum is shown for AMS~\citep[blue circles;][]{2021PhR...894....1A}, DAMPE~\citep[black triangles;][]{2017Natur.552...63D}, CALET~\citep[brown squares;][]{2023PhRvL.131s1001A}, \fermi~\citep[grey diamonds;][]{2017PhRvD..95h2007A}, and \hess~\citep[orange shaded band;][]{2009A&A...508..561A,2017PA066320,2019PhRvD..99j3022R}.}
    \label{fig:total lepton spectrum}
\end{figure}

\autoref{fig:total lepton spectrum} shows the primary CR electron flux predicted by \GP{} at the Solar location across all six source parameter combinations for the final 5\,Myr. 
The \GP{} predictions are shown by two envelopes that contain the flux curves for the central 68\% and 99\% of timesteps. 
The 99\% envelope removes significant outliers (e.g.~timesteps with Earth being contained within the injector) while demonstrating the full variability over the 5\,Myr simulation period. 

For kinetic energies below 100\,GeV the primary electron spectrum is diffuse and steady across all epochs and source parameter combinations.
At higher energies the time-dependent spectra diverge, with large fluctuations depending on the distance to the nearest electron injector.
The full modelling variation is larger than a factor of two at 1\,TeV and increases rapidly with energy.
For epochs with an electron accelerator located with $\sim$50\,pc the local primary electron flux approximately matches the injection spectrum.
For epochs with no injectors within the electron cooling distance to the Sun the local primary electron flux will be approximately equal to zero.
As the electron cooling distance via both IC and synchrotron energy losses depends on kinetic energy, we expect some energy-dependent cutoff in the primary electron flux above 100\,GeV due to propagation losses \citep[e.g.][]{2011JCAP...02..031M}.

The cut-off in the primary electron flux occurs around 300\,GeV for source parameter combinations with fewer injectors (e.g.~shorter lifetimes, L010R100, or longer times between sources, L100R500).
For source parameter combinations with larger numbers of CR sources~(e.g.~L200R100 and L100R050) there are more accelerators in the Solar neighbourhood at any time. The predicted flux is consistently high -- fluctuations due to individual sources become negligible and the flux becomes more stable over time.
The spectral cut-off in the primary electron spectrum was typically found at energies $>$10\,TeV for the source parameter combinations with larger numbers of active sources.
Across all source parameter combinations over a 5\,Myr period we found that the cut-off energy was typically around $\sim$10\,TeV.

For comparison, \autoref{fig:total lepton spectrum} also includes the combined electron and positron spectra~(with statistical and systematic uncertainties added in quadrature) for the Alpha Magnetic Spectrometer (AMS02)~\citep{2021PhR...894....1A}, the Dark Matter Particle Explorer (DAMPE)~\citep{2017Natur.552...63D}, the Calorimetric Electron Telescope (CALET)~\citep{2023PhRvL.131s1001A}, the \textit{Fermi} Large Area Telescope (\fermi)~\citep{2017PhRvD..95h2007A}, and the High-Energy Stereoscopic System (\hess)~($<$5\,TeV;~\citealt{2009A&A...508..561A}, preliminary data for energies $>$5\,TeV;~\citealt{2017PA066320,2019PhRvD..99j3022R})\footnote{During the peer-review process, updated measurements from \hess{} were reported \citep{2024PhRvL.133v1001A}. The new measurements do not impact our analyses or results.}.
The softening in the all-electron spectrum above 1\,TeV seen by the various instruments implies that there are currently no VHE lepton accelerators within the $>$1\,TeV electron cooling distance to the Solar location.
This cutoff is within our range of predictions, with the 68\% envelope across all of our model configurations lying on the upper edge of the \hess{} observations, but is generally consistent with the data.
There are two possible explanations for the predictions lying on the upper edge of measurements. Either the Earth is currently experiencing a lower-than-average CR electron flux; or, the modelling setup needs some adjustment.

Geminga is a nearby leptonic accelerator that is located at a distance of 250\,pc~\citep{2007Ap&SS.308..225F}.
Observations of the extended TeV emission from the High Altitude Water Cherenkov Experiment (HAWC) show evidence for inhomogeneous diffusion around Geminga~\citep{2017Sci...358..911A}.
Additionally, two-zone models have shown that these observations can be reproduced if CR diffusion in the nearest 30--50\,pc to the sources is much slower than diffusion further away~\citep[e.g.][]{2010ApJ...712L.153F,2018ApJ...863...30F,2019ApJ...879...91J}.
From \citet{2019ApJ...879...91J}, the cooling distance within the slow-diffusion zone (SDZ) for $\gtrsim$100\,TeV electrons was found to be $\lesssim$10\,pc.
Using the standard relation that the cooling distance is proportional to $E^{-0.25}$, we find that $\gtrsim$1\,TeV electrons to be $\lesssim$30\,pc within the SDZ.
Hence, we expect that the $\gtrsim$1\,TeV electrons from Geminga are unable to diffuse to Earth.
Including SDZs around the sources will improve the agreement between the measurements and the 68\% flux envelopes predicted by \GP.

As seen in~\citet{2019ApJ...887..250P}, the time-dependent CR flux varies around the steady-state values with large deviations~(over 10 times for the 1\,TeV CR electrons) due to the inclusion of discrete sources.
As the deviations typically increase the flux, quantifying the variability in the CRs and \grays{} via a Gaussian fit does not accurately represent the underlying distribution.
Instead, we define the containment, $S_{p}$, to quantify the factor difference from the steady-state flux to the level encompassing $p$-percent of time-dependent values -- e.g.~$S_{68}$ contains 68\% of the simulated timesteps.

We begin by calculating a list of the differences~(denoted $d_{t}$) between the steady-state flux~($J_{\mathrm{steady}}$) and the time-dependent flux values at some time $t$~($J_{t}$) in the final 5\,Myr of simulation time. These differences are calculated in logarithmic space and are given by:
\begin{equation}
    d_{t} = |\log_{10}(J_{\mathrm{steady}}) - \log_{10}(J_{t})| . \label{eq:dist from median}
\end{equation}
The differences are then sorted in ascending order. The timestep, $t$, becomes an arbitrary counting index, $i$, such that $d_{i=0}$ is the smallest difference value and $d_{i=n}$ is the largest value.
As the differences are sorted in ascending order, the containment factor is given by:
\begin{equation}
    S_{p} = \left. 10^{ d_{i} } \middle|_{i=\lfloor pn / 100 \rfloor} \right. , \label{eq:containment}
\end{equation}
where $n$ is the number of time-dependent flux values.
As $d_{i}$ is a sorted list, using the index $i=\lfloor pn / 100 \rfloor$ ensures that $p$-percent of the $n$ datapoints are within our contanment factor.
This definition of the containment factor allows us to quantify the magnitude of the asymmetric fluctuations over time at the $p$-percent level while preventing large outliers from introducing a bias towards larger values.
For example, $S_{68}=2$ states that 68\% of the time-dependent flux values across all timesteps are within a factor of two from the steady-state predictions, and $i=\lfloor 68 n / 100 \rfloor$.

The containment factor can be calculated for any time-dependent flux predicted by \GP.
For the containment factor calculations we do not restrict the analysis to times that reproduce local CR electron flux measurements. We instead calculate deviations from the steady-state values to the 68- and 95-percent levels. These containment factors will filter timesteps contaminated by outliers/local sources. Additionally, it does not limit our analyses of the CR and \graya{} variability to local electron flux measurements taken at the current~(i.e.~an arbitrary) point in time.

\begin{figure}
    \centering
    \includegraphics[width=8.2cm]{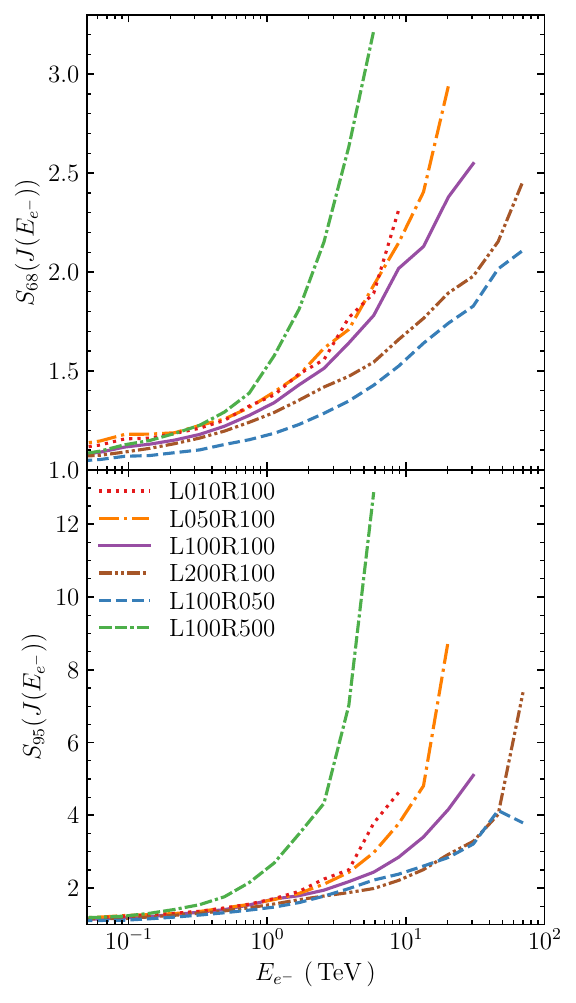}
    \caption{Containment factors for the CR electron spectra at the Solar location for the final 5\,Myr. The six source parameter combinations are distinguished by the line colours and styles. The top panel shows the $S_{68}$ containment factor and the bottom panel shows the $S_{95}$ containment factor.}
    \label{fig:variation in spectra}
\end{figure}

\autoref{fig:variation in spectra} shows the $S_{68}$ and $S_{95}$ containment factors for the CR electron flux at the Solar location for the six source parameter combinations for the final 5\,Myr of the simulation.
These results can be considered as an energy-dependent modelling uncertainty in the CR electron flux.
At 1\,TeV across all six source parameter combinations the CR electron flux has containment factors of $S_{68} \gtrsim 1.3$ and $S_{95} \gtrsim 1.5$~(i.e.~the flux varies around the steady-state values by factors of at least 1.3 and 1.5 at the 68\% and 95\% levels, respectively).
Source parameter combinations with more sources have a consistently high local electron flux as an accelerator is always in the Solar neighbourhood (within $\sim$500\,pc). Realisations with fewer sources fluctuate between high and low flux (nearby source and no nearby source, respectively), increasing the magnitude of the fluctuations.
For example, the L100R050 electron flux at 1\,TeV has a containment factor $S_{95} = 1.5$, while L100R500 is as large as $S_{95} = 2.5$.
The fluctuations also grow rapidly with energy due to the electron synchrotron and IC cooling times decreasing -- e.g.~the L100R050 containment factor increases from $S_{95} = 1.5$ at 1\,TeV to $S_{95} = 4.0$ at 50\,TeV.

\begin{figure}
    \centering
    \includegraphics[width=8.5cm]{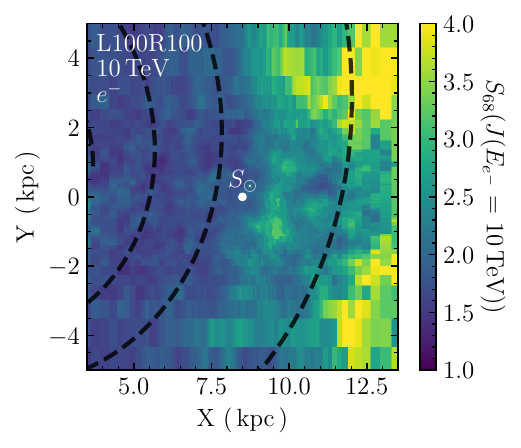}
    \caption{$S_{68}$ containment factor in the differential electron flux for the L100R100 source parameter combination at 10\,TeV across the final 5\,Myr. The Solar location is shown by $S_{\odot}$, and the containment factor is taken along the $XY$ plane~(i.e.~$Z=0$\,kpc). The spiral arms used in the source distribution are shown by the black dashed lines.}
    \label{fig:electron map}
\end{figure}

We also applied the containment factor analysis to the propagation cells in the local Galaxy, within 5\,kpc of the Solar location. \autoref{fig:electron map} shows the $S_{68}$ containment factor in the electron differential flux at 10\,TeV in the $XY$ plane for the L100R100 source parameter combination.
The results shown here agree with the residual maps found by~\citet{2019ApJ...887..250P}. However, the analysis of the fluctuations in the CR flux shown here is not dominated by individual CR sources.

For energies below 1\,TeV the electrons are able to diffuse across the inter-arm regions, leading to relatively stable fluxes over interarm distances.
The situation changes at higher energies -- above 10\,TeV the electron flux within the spiral arms is generally similar to the injected spectrum of CRs for most timesteps. The fluctuations within the arms are then relatively small.
However, the $\geq$10\,TeV electrons are unable to diffuse across the inter-arm regions due to strong synchrotron losses.
Hence, the electron flux within the inter-arm regions varies from an order of magnitude smaller than those found in the arms up to the source injection spectra.
Fluctuating between these extrema results in high magnitudes of variability in the inter-arm regions compared to those found within the spiral arms. 
From \autoref{fig:electron map}, the 10\,TeV CR electron flux in the inter-arm regions and Galactic voids can vary by up to a factor of $S_{68} \sim 4$. In contrast, the CR electron flux varies by factor of at most $S_{68} \sim 2$ within the spiral arms.
The variation in the electron flux for the other five source parameter combinations follows a similar morphology as to the L100R100 combination in \autoref{fig:electron map}.

The CR electron variation within the Galactic plane depends most strongly on the source distribution and is largely independent of the GMF strength. This behaviour may be altered by using other GMF models or by linking the spatial diffusion coefficient to the GMF strength and is beyond the scope of this work.

\section{Gamma-Ray Variability} \label{sect:gamma results}

In this paper we define the total \graya{} emission as the sum of the steady-state $\pi^{0}$-decay emission and the time-dependent IC emission.
For \graya{} energies $>$100\,GeV the Bremsstrahlung emission contributes $<$1\% to the total emission at these energies and is not included.
The \graya{} flux is calculated by performing a line-of-sight integral over the emissivities for all pixels in the all-sky maps.

\begin{figure*}
    \centering
    \includegraphics[width=16cm]{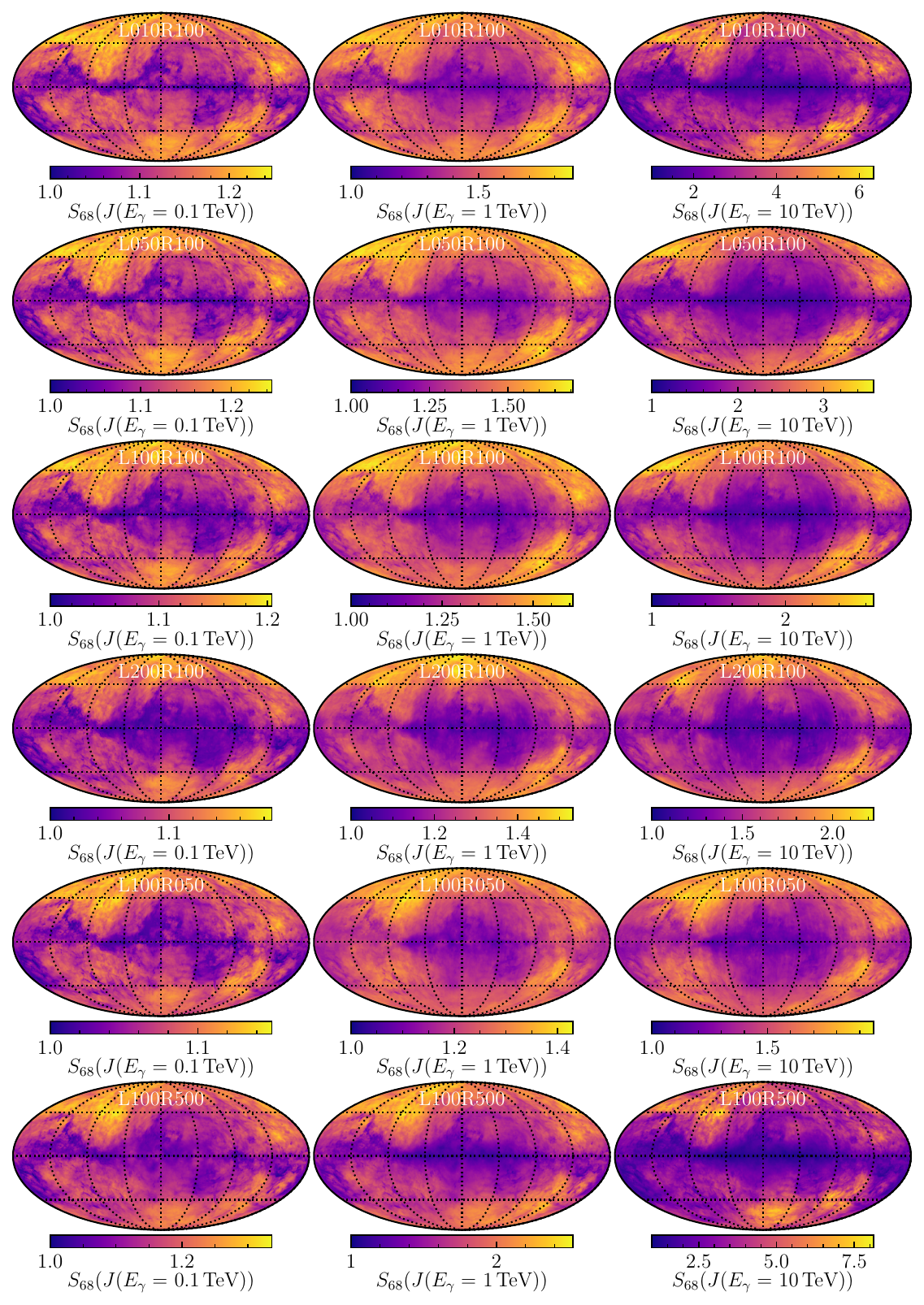}
    \caption{The total \graya{} emission $S_{68}$ containment factor skymaps across the final 5\,Myr. From left to right, the columns are: 0.1\,TeV, 1\,TeV, and 10\,TeV. From top to bottom, the rows are: L010R100, L050R100, L100R100, L200R100, L100R050, and L100R500. The colour scale changes for each skymap.}
    \label{fig:68 skymap var}
\end{figure*}

As the IC component of the total \graya{} emission is of a similar magnitude to the $\pi^{0}$-decay emission, the variability in the CR electrons throughout the MW (\autoref{fig:electron map}) will lead to fluctuations in the total \graya{} emission.
\autoref{fig:68 skymap var} shows the $S_{68}$ containment factors for the total \graya{} emission skymaps for all six source parameter combinations at 0.1\,TeV, 1\,TeV, and 10\,TeV.
The containment factor is calculated as the total variability of each pixel in the total \graya{} flux maps across the final 5\,Myr of simulation time.
The amplitude of the temporal variability of the total emission increases with energy due to both the variability in the CR electron flux growing rapidly~(see \autoref{fig:variation in spectra}) and the IC emission becoming a more dominant component of the diffuse emission above 1\,TeV~\citep{2023MNRAS.518.5036M}.

At 0.1\,TeV, all but the L100R500 source parameter combinations have a similar degree of variability, with maximum containment factors ranging from $S_{68}(J_{\mathrm{L100R050}})=1.15$ to $S_{68}(J_{\mathrm{L100R500}})=1.35$. At 1\,TeV the maximum containment factors range from $S_{68}(J_{\mathrm{L100R050}})=1.4$ to $S_{68}(J_{\mathrm{L100R500}})=2.5$, and at 10\,TeV they range from $S_{68}(J_{\mathrm{L100R050}}) \sim 1.9$ to $S_{68}(J_{\mathrm{L100R500}}) \sim 8$.
Across the sky the L100R050 and L100R500 parameter combinations are the least and most variable, respectively, for all energies. The morphology of the containment factors is consistent between the six source parameter combinations and across all energies, with line of sights with higher ISM densities exhibiting lower containment factors. The skymaps are more variable away from the Galactic plane and in the polar regions.

Across all energies and source parameter combinations, the region of the \graya{} sky with the smallest containment factors is the Galactic plane and towards the central region~($|l| < 90^{\circ}$, $|b| < 5^{\circ}$).
The increased ISM densities included in the LoS integral towards the central Galaxy average out the contributions due to individual sources. Hence, there is a reduced degree of time-dependent modelling uncertainty along the Galactic plane.
Conversely, the reduced ISM densities in the LoS integrals towards the Galactic poles and the outer Galactic regions off the plane~($|l| > 90^{\circ}$, $|b| > 20^{\circ}$) are the most variable locations in the sky across all six parameter combinations and energies.

\begin{figure}
    \centering
    \includegraphics[width=8.5cm]{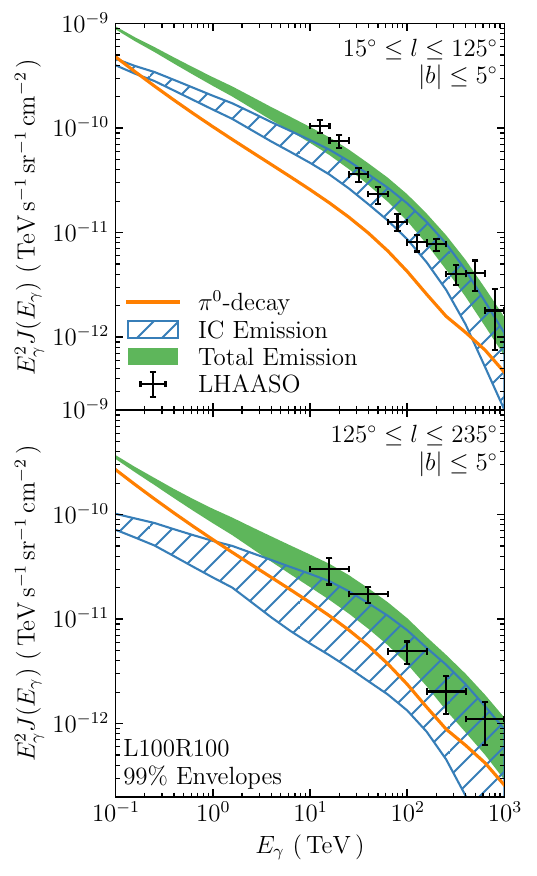}
    \caption{Differential \graya{} spectra for the LHAASO regions for the final 5\,Myr of the L100R100 source parameter combination. The steady-state $\pi^{0}$-decay emission is shown by the orange line and the 99\% envelopes are shown for the time-dependent IC emission~(blue hatched band) and the total \graya{} emission~(green shaded band). The inner and outer LHAASO regions are shown in the top and bottom panels, respectively, with the datapoints from~\citet{2023PhRvL.131o1001C}.}
    \label{fig:LHAASO individual components}
\end{figure}

The Large High Altitude Air Shower Observatory (LHAASO) collaboration recently announced a detection of the diffuse \graya{} emission along the Galactic plane above 10\,TeV~\citep{2023PhRvL.131o1001C}\footnote{During the peer-review process, updated measurements from LHAASO were reported \citep{2024arXiv241116021L}. The new measurements do not impact our analyses or results.}.
The two Galactic regions analysed by the LHAASO collaboration were the `inner' and `outer' Galactic regions, defined by $15^{\circ} \leq l \leq 125^{\circ}$, $|b| \leq 5^{\circ}$ and $125^{\circ} \leq l \leq 235^{\circ}$, $|b| \leq 5^{\circ}$, respectively.
In \autoref{fig:LHAASO individual components} we show a comparison between the LHAASO observations and the \GP{} predictions for typical Galactic source parameters (L100R100).
To ensure a fair comparison, the regions masked by the LHAASO collaboration are also masked from the \graya{} skymaps generated by \GP{} before the flux is evaluated.
For both LHAASO regions we found that the hadronic and leptonic components compete.

To highlight as much of the modelling variation as possible while not being dominated by outlier results from 1--2 timesteps, we show the 99\% envelopes for both the IC and total \graya{} emission.
We see that the 99\% envelope of the \GP{} total \graya{} predictions over the final 5\,Myr for the L100R100 source parameter combination agrees with all LHAASO flux points within uncertainties.

\begin{figure}
    \centering
    \includegraphics[width=8.5cm]{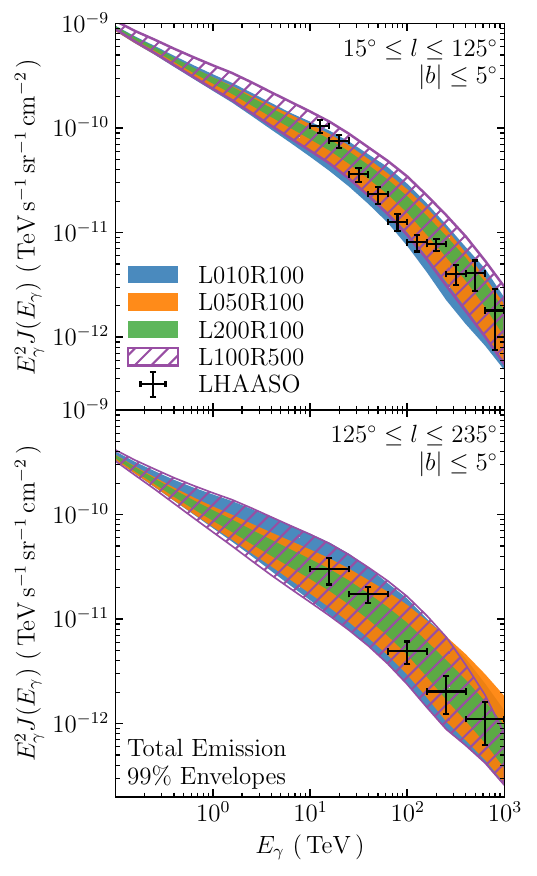}
    \caption{Differential \graya{} spectra for the LHAASO regions for the final 5\,Myr across all source parameter combinations. The 99\% envelopes are shown for the total \graya{} emission for the following source parameter combinations: L010R100~(blue shaded band), L050R100~(orange shaded band), L200R100~(green shaded band), and L100R500~(purple diagonally hatched band). The inner and outer LHAASO regions are shown in the top and bottom panels, respectively, with the datapoints from~\citet{2023PhRvL.131o1001C}.}
    \label{fig:LHAASO total}
\end{figure}

We show the total emission in the LHAASO regions for the L010R100, L050R100, L200R100, and L100R500 source parameter combinations in \autoref{fig:LHAASO total}.
As the L200R100 and L100R050 parameter combinations have a similar power of CRs injected into the MW, their \graya{} flux envelopes are approximately equal. Hence, the L100R050 flux envelope is not shown. The L100R100 envelope is shown in \autoref{fig:LHAASO individual components} and lies between the L050R100 and L200R100 flux envelopes.
Generally, the parameter combinations with fewer sources have a slightly harder IC spectrum below $\sim$100\,TeV and a softer spectrum above $\sim$100\,TeV.
However, there is significant degeneracy between the six parameter combinations within the two LHAASO regions.
All combinations adequately describe the LHAASO observations.
Hence, the diffuse emission observations from LHAASO along the Galactic plane does not currently offer further constraints for the source lifetimes or creation rates.
The full range of predictions for the six chosen source parameter combinations represents the total time-dependent modelling uncertainty in the \graya{} emission across the current best-fit models.

We also compared the predicted local electron flux and the \graya{} flux from the two LHAASO analysis regions.
No correlation was found for either LHAASO region.
Hence, both LHAASO regions probe \graya{} flux from beyond the local electron horizon.

\begin{figure}
    \centering
    \includegraphics[width=8.5cm]{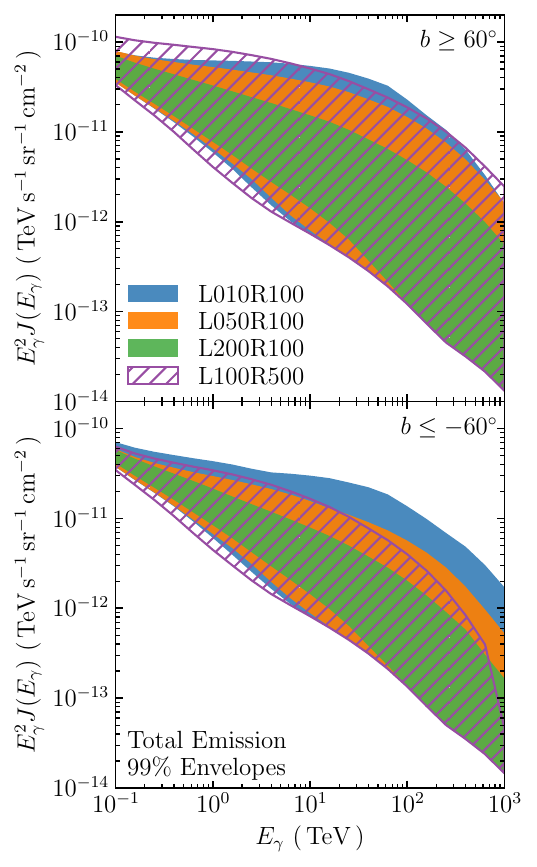}
    \caption{Differential \graya{} spectra for the polar regions for the final 5\,Myr across all source parameter combinations. The 99\% envelopes are shown for the total \graya{} emission for the following source parameter combinations: L010R100~(blue shaded band), L050R100~(orange shaded band), L200R100~(green shaded band), and L100R500~(purple diagonally hatched band). The northern pole~($b \geq 60^{\circ}$) is shown on the top panel, and the southern pole~($b \leq -60^{\circ}$) is shown on the bottom panel.}
    \label{fig:polar total}
\end{figure}

\autoref{fig:polar total} shows the \graya{} flux for the Galactic polar regions, defined by $b \geq 60^{\circ}$ and $b \leq -60^{\circ}$ for the for the north and south, respectively, over four of the six parameter combinations.
Similarly as for \autoref{fig:LHAASO total}, we found that the L200R100 and L100R050 parameter combinations almost completely overlap due to injecting similar total powers of CRs. Additionally, the L100R100 combination lies between the L050R100 and L200R100 envelopes, and so was excluded for clarity.
We found that the \graya{} flux in the two polar regions varies by over a factor of a hundred during the 5\,Myr period.
As with the LHAASO analysis regions~(\autoref{fig:LHAASO total}), the envelopes from the six source parameter combinations completely overlap. As such, it is unlikely that the source parameters can be recovered from the polar diffuse emission for a single snapshot in time.

We expect the Galactic poles to be strongly influenced by the local CR electron flux. However, restricting the analysis to only include timesteps that reproduce the local electron flux below 1\,TeV does not significantly impact the prediction envelopes for the total polar \graya{} flux.
The IC flux from the polar regions are therefore probing the electron flux beyond the local electron horizon.

Towards the polar region the isotropic diffuse \graya{} background will also contribute to the emission, and has an exponential cut off around 240\,GeV~\citep{2015ApJ...799...86A}.
Above 1\,TeV the dominant component of the \graya{} flux towards the Galactic poles is then Galactic in origin.
Towards this region the IC component of the emission is dominant from 100\,MeV to 1\,PeV due to the ISRF scale height being larger than that for the ISM gas~\citep[e.g.][]{2019ApJ...887..250P,2023MNRAS.518.5036M}.
Additionally, the flux of secondary electrons/positrons predicted by \GP{} is typically an order of magnitude lower than that of the primaries.
Therefore, the polar IC emission predicted by \GP{} is due largely to the population of primary electrons.
As the IC component from primary electrons is dominant for all source parameter combinations for the majority of the simulated epochs, and as the IC emission at these energies probe beyond the local TeV-electron horizon, the polar regions present an opportunity to constrain the current CR electron densities around the Solar neighbourhood.
LHAASO, HAWC, and SWGO may be able to observe the polar \graya{} emission in the upcoming decade as their exposures increase~\citep{2016NPPP..279..166D,2017ApJ...843...39A,2019arXiv190208429A}.

\begin{figure}
    \centering
    \includegraphics[width=8.5cm]{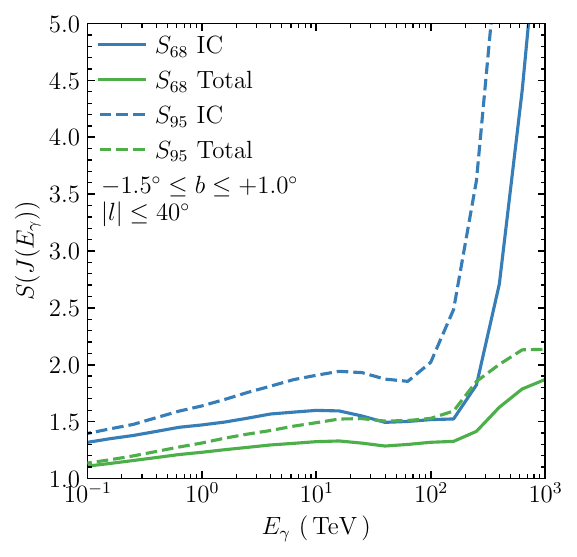}
    \caption{Containment factor in the \graya{} emission towards the GC region, defined by $|l| \leq 40^{\circ}$ and $-1.5^{\circ} \leq b \leq 1.0^{\circ}$, for the final 5\,Myr across all six source parameter combinations. The $S_{68}$ containment factor~(solid lines) and the $S_{95}$~(dashed lines) are shown for the IC \graya{} emission~(blue) and the total \graya{} emission~(green).}
    \label{fig:gray variation in GC}
\end{figure}

CTA is currently under construction and is likely to observe the diffuse emission around the Galactic centre region. With its proposed order-of-magnitude improvement to sensitivity in the TeV regime, we must consider the time-dependent modelling uncertainty towards the GC.
\autoref{fig:gray variation in GC} shows the $S_{68}$ and $S_{95}$ containment factor in the flux averaged over the HGPS analysis region (defined by $|l| \leq 40^{\circ}$ and $-1.5^{\circ} \leq b \leq 1.0^{\circ}$).
As energy increases above 1\,TeV the IC emission becomes increasingly dominant in the GC region and imparts its variability onto the total emission.
The situation changes for \graya{} energies $\geq$200\,TeV.
The time-dependent IC component becomes softer and the $\pi^{0}$-decay emission begins to contribute a larger fraction to the total emission.
However, the time-independent total flux remains harder than that of the steady-state flux up to higher energies. As the containment factor is calculated from the steady-state values, the variation grows more rapidly above 200\,TeV.

We compare the \GP{} predictions to the HGPS using a sliding window analysis.
The windows have a width of $\Delta w=15^{\circ}$ and height of $\Delta h=2.5^{\circ}$, centred at a latitude of $b_{0}=-0.25^{\circ}$ and spaced $\Delta s=1^{\circ}$ apart.
We use the HGPS flux map with an integration radius of $R_{c}=0.2^{\circ}$.
We subtract both the catalogued and unresolved source components and consider the residual HGPS emission as an estimate of the diffuse \graya{} emission integrated above 1\,TeV.
For details on the sliding window analysis and the computation of the residual HGPS emission, see~\citet{2023MNRAS.518.5036M} and references therein.
We note that the HGPS residual emission is below the survey's 5$\sigma$ sensitivity and is therefore not a detection of the diffuse emission at TeV energies.

\begin{figure}
    \centering
    \includegraphics[width=8.5cm]{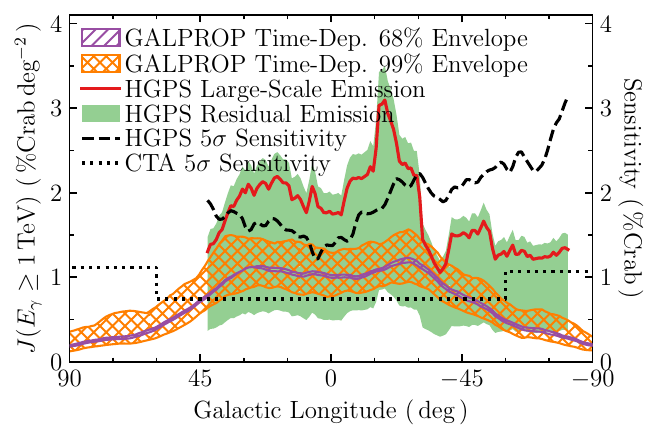}
    \caption{Longitudinal profiles integrated above 1\,TeV in units of \%Crab\,deg$^{-2}$ for the 68\% and 99\% \GP{} envelopes over the final 5\,Myr and all six source parameter combinations are shown by the purple and orange hatched bands, respectively. The HGPS profile is shown after catalogued sources are subtracted~(solid red). 
    Also shown are the results after accounting for the $\pm$30\% HGPS flux uncertainty and subtracting the contributions due to both the catalogued and unresolved sources~(green).
    The $5\sigma$ sensitivity is shown in units of \%Crab for the HGPS~(dashed black) and CTA GPS~(dotted black)
    Averaging windows were applied to all profiles~(see text).}
    \label{fig:sliding window}
\end{figure}

\autoref{fig:sliding window} shows the comparison between the HGPS longitudinal profile and the \GP{} envelopes after applying the averaging window.
The purple and orange hatched bands are the 68\% and 99\% envelopes, respectively, for the total integrated \graya{} flux above 1\,TeV across all six source parameter combinations. These bands represent the full range of time-dependent predictions.
These predictions from \GP{} for all source parameter combinations that we tested agree with the diffuse \graya{} emission estimates we calculated from the HGPS.
The temporal variability in the \graya{} emission integrated above 1\,TeV averaged along the Galactic plane is given by the containment factor $S_{95}=1.4$. The temporal variability along the Galactic plane at 1\,TeV is similar in magnitude to the modelling uncertainty over a range of ISM distributions under the steady-state approximation.
Similarly as for Figures \ref{fig:LHAASO total} and \ref{fig:polar total}, the longitudinal profiles of the diffuse TeV \graya{} emission exhibit some degeneracy between the six source parameter combinations.
Hence, observations of the diffuse TeV \graya{} emission along the Galactic plane do not currently offer any further constraints on the CR source lifetimes or creation rates.

\section{Discussion}

Our previous work~\citep{2023MNRAS.518.5036M} was the first systematic study of the modelling the variation in the TeV \graya{} emission over a grid of steady-state models.
We found that extending to higher energies required time-dependent modelling.
The work presented here is the first systematic study of the time-dependent modelling variation that arises from placing CR sources stochastically throughout the MW.

It has been well known for decades that the CR electron spectrum measured at Earth (see \autoref{fig:total lepton spectrum}) is not representative of the Galactic spectrum~\citep[e.g.][]{1997JPhG...23.1765P}.
If there are no electron accelerators within $\sim$200\,pc of the Earth then a break in the local spectrum at TeV energies is expected~\citep[e.g.][]{2021PhRvD.104l3029E,2023PhRvD.108j3015S} independent of any cutoff in the injection spectrum.
Further Monte~Carlo results from~\citet{2018JCAP...11..045M} have shown that any potential break in the local CR electron spectrum can arise naturally from a population of discrete CR sources.
For this reason we used an electron injection spectrum that is fitted in the GeV energy range that reproduces the local CR observations only up to 1\,TeV. We do not apply any spectral alterations~(e.g.~by introducing a cut off to the injection spectrum in the TeV regime) as many PWNe have recently shown some evidence of electron acceleration up to energies of 1--10\,PeV \citep[e.g.][]{2021Natur.594...33C,2021Sci...373..425L,2022ApJ...930..148B,2024NatAs...8..628Y} for at least the first 15\,kyr of their lives~\citep{2024RAA....24g5016L}.
Our results are able to reproduce a TeV cutoff in the local electron spectrum for timesteps with no nearby sources.
From \autoref{fig:variation in spectra} we found that the modelling uncertainty at the 68\% level in the CR electron flux is larger than $\pm$50\% at 10\,TeV. At the 95\% level the flux varies by over a factor of four.
As the electron flux above 1\,TeV is so strongly coupled to the locations of nearby accelerators, CR propagation codes (such as \GP) must consider their placement of discrete sources in this energy regime.
The spectral shape of the CR electron observations above 1\,TeV should be reproduced by considering the placement of the CR sources.
Including effects such as slow-diffusion zones around sources may be required to improve the match between observations and the 68\% envelope predictions.
Other alterations to the models, e.g.~altering the source injection spectrum, is not required to reproduce CR observations.

\subsection{Resolved, unresolved, and diffuse emission}

We consider CRs to be local to their sites of acceleration while their trajectories are governed by the dynamics of the source~(e.g.~magnetic fields, turbulence, etc.). Any \grays{} emitted by these CRs are defined as `source' or `local' emission.
For the \graya{} source to be resolved it must be luminous enough to be detected confidently above the diffuse background emission.
However, even if the source cannot be confidently detected, the \graya{} flux local to it will still contribute to the observations.
We refer to these \grays{} as unresolved source emission.
Diffuse emission is then what remains -- \grays{} from CRs which are not governed by the dynamics of their acceleration site.
The large-scale residual \graya{} emission, found after subtracting all resolved source emission and a relative contribution from unresolved sources, is assumed to be the diffuse emission.
Due to the uncertain nature of unresolved emission, it is widely accepted that there will be some contamination in the observations.

For this paper, the hadronic component of the emission is diffuse by construction~(see~\autoref{sect:hadron results}).
Our modelled leptonic component the \grays{} can be either diffuse or source emission.
All IC emission from secondary leptons (i.e.~leptons created via hadronic interactions) is diffuse by construction of our models. For the primary electrons, i.e.~those injected by sources, only the IC emission from once they have diffused far from their sources is diffuse.
There will be some threshold energy in the range of 1--100\,TeV above which the primary electrons cool before they can escape the source environments.
Most of the IC emission that remains in the residual emission above $\sim$10\,TeV is a spatially unresolved component.

LHAASO has detected the diffuse emission above 10\,TeV~\citep{2023PhRvL.131o1001C}, with their estimates lying above the hadronic, diffuse component of our steady-state predictions~(e.g.~\autoref{fig:LHAASO total}).
There are multiple potential explanations. There could be an unknown additional diffuse hadronic component, new physics, or there could be some contamination from sources.
We found that the LHAASO observations can be reproduced by current CR propagation models by considering a steady-state hadronic component with a time-dependent leptonic component.
Hence, no additional hadronic components (or other alterations to the models) are required to reproduce the observations\footnote{The \argo{} TeV diffuse emission results~\citep{2015ApJ...806...20B} are also reproduced by \GP. We do not show comparisons as the \argo{} analyses use different regions to LHAASO.}.
Additional constrains on the hadronic emission can be obtained from the recent IceCube observations~\citep{2023Sci...380.1338I}.
The neutrino predictions from \GP{} are in full agreement with the IceCube observations, and will be shown in a forthcoming paper.
Therefore, the LHAASO excess can be explained by \GP{} under the chosen set of models without any additional hadronic components.

Our comparisons to LHAASO imply that \graya{} sources contribute approximately half of the measured diffuse emission in the `inner' region.
This source component could be from unresolved sources, agreeing with the results from~\citet{2024NatAs...8..628Y} which found that unresolved PWNe are able to make a considerable contribution to the measured diffuse \graya{} emission observed by LHAASO.
The source component could also have some contribution from possible residual emission from resolved sources that were not fully masked~\citep{2024arXiv240715474C}, though we are not able to comment on the fraction of contamination versus unresolved flux.

We should also note that for kinetic energies $\gg$1\,PeV the CR transport starts to transition from the diffusive to the ballistic regime.
As \GP{} reproduces the LHAASO emission, use of the diffusion approximation remains appropriate for calculations of the \graya{} emission up to 1\,PeV.
For higher energies, and as observations improve, the use of the diffusion approximation may need to be reevaluated.

\subsection{Variability of the interstellar emissions}

We quantified the temporal variability of the \graya{} flux by using the containment factors given in \autoref{eq:containment}. This quantification of the variability is shown across the whole sky in \autoref{fig:68 skymap var} and for the GC region in \autoref{fig:gray variation in GC}.
The variability of the total \graya{} emission depends on both the energy of the \grays{} and the amount of the ISM in the LoS integral, with denser areas of the ISM being less variable over time.
For the chosen model components, the polar regions and the outer Galaxy are the most variable regions in the sky.

As the CR source lifetimes and creation rates are not precisely known, we simulated over a range of reasonable parameter combinations~(see \autoref{tab:source pars}).
We find that if the number of sources per unit time increases, the temporal variability decreases.
The diffuse \graya{} emission has a strong dependence on the CR source parameters.
However, there is significant degeneracy between the predictions, with all of the tested combinations reproducing the diffuse emission observations.
The properties of the CR sources plays some factor in the spectrum and brightness of the diffuse emission but are not necessarily recoverable from either on- or off-plane observations.
The source properties and placement impact the \graya{} morphology nearby to the sites of acceleration. However, the all-sky morphology over long periods of time is largely identical between the source parameters, with only the amplitude of the variability increasing for models with fewer CR sources~(as seen in \autoref{fig:68 skymap var}).
None of the source parameter combinations that we tested can be excluded by current measurements of the diffuse \graya{} sky.

For on-plane regions the steady-state $\pi^{0}$-decay is a significant component of the diffuse \graya{} emission.
Additionally, as the CR electrons have a theoretical maximum energy on the order of a few PeV~\citep[e.g.][and references therein]{2021Sci...373..425L}, and as the IC scattering cross section becomes KN suppressed, the leptonic emission will rapidly become sub-dominant to the $\pi^{0}$-decay above a few PeV.
Therefore, considering a time-dependent solution to the transport equation is a modelling consideration only for the \graya{} emission in the energy range of $\sim$1\,TeV to $\sim$1\,PeV.

Towards the GC our estimates of the \graya{} variability~(\autoref{fig:gray variation in GC}) can only be considered as lower limits as we do not model the GC as a separate variable CR accelerator.
Additionally, we do not model the CR sources with injection spectra that evolve over time~\citep[e.g.][]{1996A&A...309..917A,2007ApJ...665L.131G,2010PASJ...62.1127C} and we do not consider two-zone diffusion around sources~\citep[e.g.][]{2010ApJ...712L.153F, 2018ApJ...863...30F,2019ApJ...879...91J}.
While \GP{} is currently able to model both of these effects, we do not use that functionality in this paper.
We also do not model lower-energy CRs being trapped within the source volume for longer than the higher-energy particles~\citep[e.g.][]{2009MNRAS.396.1629G,2021MNRAS.503.3522M}.
Including these three effects increases the complexity of the simulations and is left for future studies. Trapping lower-energy CRs and utilising injection spectra that evolve over time are expected to increase the variability in the \graya{} sky.
Utilising two-zone models will also introduce a spectral cutoff to the local CR electron spectrum~(\autoref{fig:total lepton spectrum}) more often, reducing the predicted envelope across all models.

\subsection{Dependence on the source distribution}

Our previous results in~\citet{2023MNRAS.518.5036M} showed that the total TeV \graya{} emission did not strongly depend on the chosen ISRF.
Additionally, the structure of the GMF was observable in the IC emission above 10\,TeV but had a minor impact on the total emission.
We found that the source distribution was the Galactic model parameter that had the largest impact on the total \graya{} emission. Towards the GC at 1\,TeV the source distribution impacted the total emission by $\sim$20\%. 
We were unable to perform time-dependent simulations over the same five CR source distributions used in \citet{2023MNRAS.518.5036M} (SA0, SA25, SA50, SA75, and SA100) in this work due to the required computational resources.
Simulations were performed with the SA100 source distribution for a small-scale comparison to the SA50 model. The results are summarised here.

The SA source distributions are normalised such that the CR injection around the Solar neighbourhood is consistent between them~\citep{2017ApJ...846...67P}. For the time-dependent simulations, this normalisation condition results in the probability of a CR source being placed within a few hundred parsecs of the Solar location being approximately equal across the source distributions.
The 99\% envelopes for the CR electron flux at the Solar location was found to be consistent between the SA50 and SA100 distributions.

The variability in the total \graya{} flux was also similar between the SA50 and SA100 source distributions.
Below 1\,TeV the \graya{} flux and containment factors are broadly similar between SA50 and SA100, with differences being $\lesssim$5\% within the Galactic plane.
Along the spiral-arm tangents the differences are larger, with the upper bounds of the longitudinal profiles typically being 10\% higher for the SA100 distribution at 1\,TeV. However, the minimum bound of the total \graya{} flux along the spiral arm tangents is approximately equal between the two source distributions. Hence, the SA100 distribution is $\sim$5\% more variable along the spiral-arm tangents.

Across the entire simulated energy range the impact of the source distribution is minor compared to the variability that is introduced from the time-dependent solution.
For example, at 1\,TeV the difference between the source distributions along the spiral-arm tangents is $\sim$20\%. Along the spiral-arm tangents the variability introduced by the time-dependent solution can be as large as $\sim$60\%.

\section{Summary}

We have performed a systematic comparison between observations and the time-dependent CR and \graya{} fluxes predicted by \GP.
The time-dependent models exhibit large fluctuations in both the CR electron and total \graya{} fluxes, with the magnitude of the variations depending on the underlying source parameters such as their injection fluxes, lifetimes, and their creation rates.
To quantify the full range of time-dependent uncertainties we simulated over six source parameter combinations with lifetimes 10--100\,kyr and creation intervals 50--500\,yr~(\autoref{tab:source pars}).
As expected, the source parameter combinations with fewer sources and shorter-lived sources have the largest fluctuations for both the CR and \graya{} fluxes.
While the source parameters influence the spectral shape and brightness of the diffuse emission, we found the resulting CR and \graya{} fluxes exhibit significant degeneracy with one another. On the basis of a single-class source distribution, the underlying source parameters are not recoverable from observations of either the local CR flux or diffuse \graya{} emission for any single snapshot in time.
However, we have assumed that there is only one `average' source class.
The source parameters may be more easily extracted if using multiple source classes~(e.g.~SNRs, PWNe, stellar clusters, etc.), with each distributed uniquely, and is left for future work.

The magnitude of the fluctuations in the local CR electron flux above 1\,TeV can be as large a factor of ten over a period of 5\,Myr.
The fluctuations in the leptonic \graya{} emission are typically as large as factors of 1.2 to 2 depending on the region, representing a significant modelling uncertainty for the total emission.
At 1\,TeV the fluctuations in the \graya{} emission along the Galactic plane found by our time-dependent modelling is similar in magnitude to the variations over a grid of steady-state models found by~\citet{2023MNRAS.518.5036M}.
For the outer Galactic regions the time-dependent fluctuations for the total emission can be as large as a factor of ten.
For the Galactic polar regions the total \graya{} emission can fluctuate by over a factor of 100 over a 5\,Myr period.
Additionally, as the \graya{} emission towards the Galactic poles in the TeV--PeV regime is purely Galactic in origin and dominated by the IC component, future observations will be able to constrain the 1--100\,TeV CR electron flux outside of the Solar neighbourhood.

Due to the KN suppression of the IC scattering cross section for the ISRF and CMB photon fields, the leptonic fraction of the simulated \graya{} emission begins to decrease above 100\,TeV.
As the hadronic component is steady in the VHE regime, utilising a time-dependent solution is only required for the computation of the leptonic component for \graya{} energies between $\sim$1\,TeV to $\sim$1\,PeV.

Our \GP{} predictions of the diffuse \graya{} emission for all timesteps agree with the lower limits inferred from the HGPS~\citep{2018A&A...612A...1H} after accounting for both the catalogued and unresolved source components.
We also compared the time-dependent predictions to the diffuse emission observations by LHAASO up to 1\,PeV~\citep{2023PhRvL.131o1001C}. After the same exclusion regions used in the LHAASO analysis were masked from our predictions we found the models and observations are in agreement within their respective uncertainties.
Hence, \GP{} has now been shown to successfully reproduce the Galactic diffuse \graya{} emission from energies below 100\,GeV~\citep[e.g.][]{2016ApJS..223...26A,2016ApJS..224....8A,2016ApJ...819...44A} up to 1\,PeV.
As the LHAASO emission is reproduced, additional hadronic components, such as increasing the ISM gas density, is required.
The most likely explanation for the excess \graya{} emission observed in the 10\,TeV to 1\,PeV energy range is then contamination from unresolved sources.
We have found that the diffusion approximation is able to reproduce current observations for \graya{} energies up to 1\,PeV.

\begin{acknowledgments}
This research was partly supported by an Australian Government Research Training Program Scholarship. \GP{} development is partially funded via NASA grants 80NSSC22K0477, 80NSSC22K0718, and 80NSSC23K0169. Some of the results in this paper have been derived using the \verb|HEALPix|~\citep{2005ApJ...622..759G} and \verb|Astropy|~\citep{013A&A...558A..33A, 2018AJ....156..123A} packages. This work was supported with supercomputing resources provided by the Phoenix HPC service at the University of Adelaide, and we want to thank F.~Voisin in particular.
\end{acknowledgments}


\appendix
\section{Time-Dependent Hadron Results} \label{sect:hadron results}

As discussed in \autoref{ssect:t-dep parameters}, propagating hadrons is considerably more computationally expensive compared to leptons for time-dependent simulations. Both hydrogen and helium must be propagated for accurate computation of the \graya{} skymaps, and comparisons to the B/C ratio requires the inclusion of elements up to and including iron~(though including up to silicon is typically sufficient). Additionally, while the CR electrons only require approximately 100\,Myr to reach their steady-state fluxes, the hadronic species require 450\,Myr of simulation time~\citep{2019ApJ...887..250P}.
These computations are unnecessary as the CR hadrons have long residence times~($\sim$1\,Gyr) and multi-kiloparsec cooling distances. Hence, we can expect the hadronic species, and the related hadronic \graya{} emissions, to be steady over time.

Here we use the same model setup as for the time-dependent leptonic runs discussed in \autoref{ssect:t-dep parameters}, with the diffusion and spectral parameters given in the supplied online material. We propagate CRs with nuclear charges $Z \leq 14$~(i.e.~up to and including silicon) with ten bins per decade ranging from 1\,GeV\,nuc$^{-1}$ to 1\,PeV\,nuc$^{-1}$, allowing accurate computation of the secondary boron fluxes up to $\sim$30\,TeV\,nuc$^{-1}$.
The source lifetime was set to 100\,kyr and the source creation rate was set to one source per century, i.e.~we use the L100R100 source parameter combination. The hadronic species were propagated for an initial 500\,Myr to allow all CR species at all energies to reach their steady-state values. The CRs were then propagated for an additional 5\,Myr with a timestep $\Delta t = 5$\,kyr and an output frequency of once every 25\,kyr. All analyses in this appendix are performed on the final 5\,Myr.

\begin{figure}
    \centering
    \includegraphics[width=8.cm]{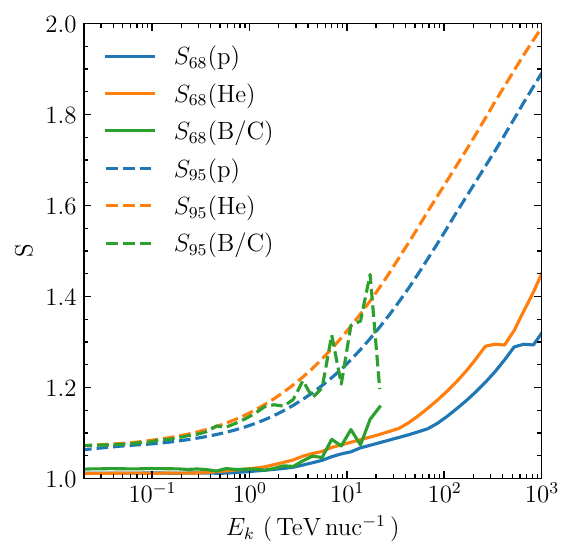}
    \caption{Variation factors for the hadronic spectra at the Solar location for the final 5\,Myr for the L100R100 source parameter combination. The \GP{} containment factors are calculated over a 5\,Myr period, and are shown by: $S_{68}$~(solid lines), $S_{95}$~(dashed lines), protons~(blue), helium~(orange), and the B/C ratio~(green).}
    \label{fig:CR variation}
\end{figure}

As is done for the local electron and \graya{} fluxes, we use the containment factor~(\autoref{eq:containment}) to quantify the variations in the CR hadrons. \autoref{fig:CR variation} shows the containment factors for the proton, helium, and B/C ratios up to 1\,PeV. The variation in the three spectra are similar in magnitude to one another. The hadronic spectra are effectively constant below 1\,TeV. Above 1\,TeV the CR fluxes are generally higher than the steady-state values. The containment factors reach up to $S_{68}=1.05$ and $S_{95}=1.20$ at 10\,TeV. This stability is in contrast to the electron variation found in \autoref{fig:variation in spectra}, which shows that the local electron flux at 10\,TeV varies by the containment factors of $S_{68}=2.1$ and $S_{95}=3$.
Furthermore, the \graya{} flux is generally more stable than the CR flux due to the LoS integral averaging over small variations.
Hence, any time-dependent variations in the total \graya{} emissions will be dominated by variations in the leptonic \graya{} fluxes. Using the steady-state assumption for the hadronic \graya{} flux is valid until observational uncertainties of the various \graya{} observations are reduced.

\bibliography{ref.bib}{}
\bibliographystyle{aasjournal}



\end{document}